\documentclass[aps,onecolumn,10pt,final, showpacs]{revtex4}
\usepackage{amsmath}
\usepackage{amssymb}
\usepackage[russian]{babel}
\usepackage{graphics}
\usepackage{graphicx}

\def\nhh{\hspace*{-0.3em}}
\def\cm{\hspace*{1cm}}

\def\lal{&& {}\nhh}

\def\beq{\begin{equation}}
\def\eeq{\end{equation}}
\def\bear{\begin{eqnarray}}
\def\bearr{\begin{eqnarray} \lal}
\def\ear{\end{eqnarray}}
\def\earn{\nonumber \end{eqnarray}}
\def\nn{\nonumber\\ {}}

\def\eql{&\! = &\!}

\def\mn{_{\mu\nu}}

\def\cV{{\cal V}}

\def\kappa{\varkappa}
\def\d{\partial}
\def\e{{\,\rm e}}

\def\sign{\mathop{\rm sign}\nolimits}
\def\1s{{\rm{I}_s}}
\def\1t{\rm{I}_t}
\def\1{\rm I}

\def\Half{{\frac{1}{2}}}

\def\wt{\widetilde}
\def\tg{{\wt g}}

\begin{document}

\title{On the Origin of Gauge Symmetries and Fundamental Constants}

\author{S.G. Rubin}
\affiliation{National Research Nuclear University ''MEPhI'' , Kashirskoe sh. 31, Moscow, 115409 Russia \\
E-mail: sergeirubin@list.ru}

\begin{abstract}
{A statistical mechanism is proposed for symmetrization of an extra space. The conditions and rate
of attainment of a symmetric configuration and, as a consequence, the appearance of gauge invariance in
low-energy physics is discussed. It is shown that, under some conditions, this situation occurs only after completion of the inflationary stage. The dependencies of the constants  $\hbar$ and G on the geometry of the extra space
and the initial parameters of the Lagrangian of the gravitational field with higher derivatives are analyzed.}

\pacs{04.50.-h; 04.05.Cd; 11.30.Ly}

\end{abstract}
\maketitle

\section{INTRODUCTION}

In spite of the advances made in fundamental physics, the existence of a large number of theories and
approaches suggests that there are problems in
describing observational and experimental data. The
idea of compact extra dimensions allows one to effectively explain a considerable number of phenomena
and indicates the direction of further development in
the theory despite the absence of direct experimental
confirmation of its existence.

This paper investigates the problems associated
with the emergence of gauge symmetries and fundamental constants at the early stage of the evolution of
the Universe using the approach proposed in our previous works
в \cite{BR06,Rubin08}. This approach is based on the assumption that extra dimensions exist forming a
compact space with the properties determining the observed low-energy physics. The solution to the first
problem within the Kaluza - Klein approach is well known: gauge symmetries arise as a consequence of
the corresponding symmetries of the extra space (see, for example, \cite{Blagojevic}). Therefore, the problem is reduced
to justification of the choice of symmetric spaces
among a set of extra spaces with an arbitrary geometry.
More precisely, we assume that, at some instant of
time on the Planck scale, the four-dimensional Riemannian space
$M_4$ \cite{HH,V2,Wilt} arises as a result of quantum fluctuations. Simultaneously, there arises a compact
extra $d$ - dimensional space  $M_d$. The set of possible
geometries of the extra space $M_d$ is at least a continuum. However, there exists a geometry with a high
degree of symmetry, because the existence of gauge
symmetries is beyond question. The probability of this
event is negligible, and, hence, there should exist a
mechanism of selection that separates appropriate geometries. One of the possible variants based on statistical considerations will be analyzed below.

The second problem discussed in our work is related to the fundamentality of the parameters $\hbar , G$ \cite{Duff} 7] and consists in the following. The geometric approach to the theory implies the presence of only one scale of the dimension of length $\ell$. The parameters of the initial Lagrangian constructed only from the metric tensor are proportional to $\ell ^n, n$, where n is an integer number. The question arises as to the instant of time at which the Planck and gravitational constants appear as independent fundamental parameters and as to their relation to the initial parameters of the theory.
The answer to this question for the gravitational constant is well known and, as a rule, is represented in the form
\begin{equation}\label{MP}
M_P ^2 = m_D ^{D-2} V_d .
\end{equation}
Here, $M_P$ is the Planck mass, $m_D$ is the parameter with the dimension of mass or, more exactly, inverse length,
and $V_d <\infty$ is the volume of the extra space $M_d$. The similar problem of the fundamentality of the Planck
constant  $\hbar$  is less well understood, even though interesting works have appeared in this direction. In particular, Volovik [8] discussed the place of the Planck constant in modern theory, analyzed different variants of
including this constant in field equations, and considered the consequences of the hypothetical spatial
dependence of the Planck constant. However, specific implementation of the aforementioned hypothetical
variants was not given.

In this paper, we propose the mechanism of emergence of the Planck constant together with the gravitational constant without detailed discussion of the possibilities this entails. As a result, we can reveal the
relations of the constants $\hbar$ and $G$ determined from low-energy experiments to the parameters of the initial Lagrangian describing multidimensional gravity with higher derivatives.

As a continuation of the ideology developed in  \cite{Rubin08},
we assume that spaces with an appropriate geometry arise in a spacetime foam with some, even though small,
probability. We are interested in spaces with a geometry of the direct product
\begin{equation}\label{MD}
M_{D}=M_{D_1}\otimes M_{d_1};\quad D_1 \geq 4;\quad d_1 \geq 2
\end{equation}
and the relationship of the volumes $V(D_1)>>V(d_1)$. The transitions with a change in the geometry are conveniently described in terms of the path integral technique
\cite{HH,V2}. For this purpose, the superspace $\mathcal M_D =(M_D ; g_{ij})$ is defined as a set of metrics $g_{ij}$ in the space $M_D$ to within diffeomorphisms. On a spacelike section
$\Sigma$ we introduce the metric $h_{ij}$ (for details, see \cite{Wilt}) and
define the space of all Riemannian $(D-1)$ metrics in
the form
\begin{equation*}
Riem(\Sigma ) = \{ h_{ij} (x)\left| {x \in \Sigma } \right.\}
\end{equation*}
The transition amplitude from the section $\Sigma _{in}$ to the section $\Sigma _{f}$ is the integral over all geometries allowable by the boundary conditions:
\begin{equation}
A_{f,in}=\left\langle {h_{f},\Sigma _{f}|{h_{in},\Sigma
}_{in}}\right\rangle =\int_{h_{in}}^{h_{f}}Dg\exp [iS(g)] .
\label{Amplitude}
\end{equation}

The absence of the Planck constant in the exponential
function is a result of choosing the appropriate units of
measurement. Nonetheless, it will be shown below
that the Planck constant naturally emerges after the
definition of the dimensional units simultaneously
with gravitational constant $G$. However, the statement
on the unification of gravity and quantum theory
would be premature. The essence of quantum
mechanics is based on the rule of summation of the
transition amplitudes  (\ref{Amplitude}), which is postulated originally.

\section{WHY IS THE EXTRA SPACE SYMMETRIC?}

The transition amplitude (\ref{Amplitude}), generating a four-dimensional space, as a rule, has been calculated
under the assumption of a special form of the metric $h_f$ on a hypersurface $\Sigma _{f}$ with an interval of the type \cite{V1,Wilt}
\begin{eqnarray}\label{metric}
&&ds^{2}=\sigma ^{2}\left[ N(t)^{2}dt^{2}-a(t)^{2}d\Omega
_{3}^{2}\right] ,\\
&&\sigma ^{2}=\frac{1}{12\pi ^{2}M^{2}_{Pl}}. \nonumber
\end{eqnarray}
In this case, it is implied that the space with a metric
that weakly differs from the metric written above
asymptotically tends to (\ref{metric}) in the course of cosmological expansion. If the transition amplitude (\ref{Amplitude}) describes the birth of a space of type
(\ref{MD}), where $D_1=4$, no preferred geometry exists for the subspace $M_{d_1}$. In
this section, we propose the mechanism of symmetrization of compact spaces.

Let us postulate that the evolution of any closed
system is accompanied by an increase in its entropy. It should be noted that the entropy of a subsystem can
decrease. In particular, the Hawking evaporation of a
black hole decreases its entropy but increases the
entropy of the Universe as a whole. Below, we will consider a space with symmetry (\ref{MD}) under the assumption
that the volume of the subspace is considerably larger than the volume of the compact subspace $M_{d_1}$.

\subsection{Entropy of the Compact Space}

First and foremost, we will demonstrate that the entropy of a compact space reaches a minimum on a
class of maximally symmetric spaces. The initial definition of the entropy in our work coincides with the
known definition of Boltzmann, who related the entropy to the number $\Omega$ of states of a system: $s =k_B \ln\Omega $. The discussion of other possibilities can be found, for example, in \cite{Brun,Katok}. The notion of the
number of states is correctly defined at the quantum level, when the set of energy levels and the degree of
their degeneracy are known. However, the quantization of the space  $M_{d}$ means the quantization of a gravitational field, which by itself remains an unsolved problem. In this respect, we will restrict ourselves to
the calculation of the number of states from the classical viewpoint. This is sufficient because we are interested in relative quantities (see also  \cite{Starkman00}).

The entropy of an arbitrary compact space $M_{d}$, in
which fields of matter are absent is a functional $s[G]$ of
the metric tensor $G$. The number of states is determined by an observer outside the system. It is intuitively clear that the higher the symmetry of an object
(in our case, the compact space), the smaller the statistical weight of the object. Let us prove this statement. In order to calculate the statistical weight $\Omega$, the space $M_{d}$ is embedded in the space $R^N$, which can be
done if the space $M_{d}$ is sufficiently smooth and $N$ is sufficiently large \cite{Dubrovin}.
It is assumed that the external observer is located in the space $R^N$. Each point of the
space $M_{d}$ is described by the internal coordinate $y\in M_d$ and the external coordinate $x\in R^N$. We choose a
specific point $P\in M_{d}$ and fix its coordinate $x_P\in R^N$. Then, we choose the set of basis vectors $e_k , k=1,2,...,N$ in the space $R^N$. We require that all $d$ tangent linearly independent vectors at the point $P$, $e^{(P)}_a , a=1,2,...,d$, which form the coordinate basis in the space $M_d$,
should be included in this set. With the use of this basis, we calculate the components of the metric tensor $G_{ab}^{(P)}(x_P )$ of the space $M_{d}$. We fix the tangent space $T(P)$,
spanned by the vectors $e^{(P)}_a$.

Then we choose the second point $Q\in M_{d}$, with the coordinates $x_Q\in R^N$
and the tangent space $T(Q)$. By moving along the curve $l_{PQ}\subset M_{d}$, the point $Q$ is displaced to the point with the coordinates $x_P$ so that the tangent space $T(Q)$ coincides with the tangent space $T(P)$.

The new components of the metric tensor $G_{ab}^{(Q)}(x_P)$, of the space $M_{d}$ are calculated at the same point     $x_P \in R^N$. If the metric tensors do not coincide with each other $G_{ab}^{(Q)}(x_P )\neq G_{ab}^{(P)}(x_P )$, the observer in the space $R^N$ will fix a new ''microstate'' and increase the statistical
weight of the space $M_{d}$ by unity. When the equality
\begin{equation}\label{Li}
G_{ab}^{(Q)}(x_P ) = G_{ab}^{(P)}(x_P ),
\end{equation}
is valid, the number of microstates remains
unchanged. It should be noted that condition (\ref{Li}) corresponds to the condition for the existence of the Killing vector along the curve $PQ$. Therefore, the presence of Killing vectors decreases the statistical weight of the compact space. Maximally symmetric spaces have a minimum entropy.

\subsection{Decay of Excitations of the Compact Space}

The compact subspace $M_{d_1}$ can be considered as a subsystem of the space $M_D$, (see relationships (2)). Now, we show that, if the volumes of the subspaces satisfy the inequality $V_{D_1}>>V_{d_1}$, there is an entropy flow from the subspace $M_{d_1}$ to the subspace $M_{D_1}$ the
entropy of the subspace $M_{d_1}$ tends to a minimum, and its geometry tends to a maximally symmetric geometry. Furthermore, the entropy of the entire system, i.e., the space $M_{D}$ increases.

As a rule, the entropy of a closed system changes
with conservation of the system energy. Since the
determination of the gravitational energy depends on
the space topology, we will restrict our consideration to transitions without changes in it. In \cite{Zhuk,our}, it was
demonstrated that, in the framework of gravity nonlinear in the Ricci scalar, at least local minima of the energy density exist. In this case, there is a set of energy levels and each geometry of the extra space can be represented in the form of an expansion in eigenfunctions of the d'Alembert operator of the background metric. In terms of the Kaluza - Klein theory, the eigenvalues of this operator contribute to the mass of excitations, which are interpreted as particles in the space $M_4$. For a compact space with the geometry of a circle of radius $r$,
the mass of the lightest particle is $m_1 =1/r$ and it is experimentally limited from below by
a value of several TeV. It is evident that its decay into
light particles propagating in our space should be
accompanied by an increase in the entropy. In the process, the geometry of the extra space relaxes to a state that has a minimum entropy and is characterized by the absence of excitations.

As an illustration, we consider a space of type (\ref{MD}), where
$M_{D_1}=M_4\otimes M_{d_2}, M_{d_2}= S_1 , M_{d_1}= S_1$ with the metric
$g_{MN}$, which differs from the diagonal metric
$\eta_{MN}=diag(1,-1,-1,-1,-r_1,-r_2)$, only slightly, so that
$g_{MN}=\eta_{MN}+h_{MN}(x, y_1 , y_2)$. The size of one of the extra spaces is considerably larger than the size of the other extra space: $r_2 = \beta r_1
,\quad \beta >>1$. The dynamic equation for the field $h_{MN}$ can be written in the form \cite{Blagojevic}
$$\eta ^{AB}\partial_A \partial_B h_{MN}(x, y_1 , y_2)=0.$$
By substituting the field $h$ in the form of the expansion in eigenfunctions of the d'Alembert operators $Y_n (y_1 ) , Y_n ( y_2 )$ оof both subspaces (in our case, circles)
$$h_{MN}(x, y_1 , y_2)=\sum_{n_1 , n_2}h^{(n_1 ,n_2)}_{MN}(x)Y_{n_1} (y_1 ),Y_{n_2}
( y_2 ),$$ we obtain the equation for the components
\begin{equation}\label{exitation}
\left( \square _x +\frac{n_1 ^2}{r_1 ^2}+\frac{n_2 ^2}{r_2
^2}\right) h^{(n_1 ,n_2)}_{MN}(x) =0.
\end{equation}
The microstates of the subspaces $M_{d_1}$ and $M_{d_2}$ are characterized by the integer numbers $n_1$ и $n_2$. The first excited state of the subspace $M_{d_1}$ has a mass $m_1 = 1/r_1$.
The decay of this state into two excited states of
the subspace
$M_{d_2}$ with masses $m_2 =n_2/r_2$ and $m'_2 =n'_2/r_2$, $n_2 , n'_2
=0,1,2,...,[\beta]$ and zero momenta
occurs with energy conservation,
\begin{equation}\label{mcons}
m_1 =m_2 +m'_2
\end{equation}
and an increase in the entropy of the entire system.
This increase is associated with the fact that the number of different microstates of the space $M_{d_2}$, that satisfy condition (\ref{mcons}) is of the order of$\Omega_2 =\beta
/2 \sim r_2 /r_1 $. In this case, the geometry of the subspace $M_{d_1}$ tends to a maximally symmetric geometry, because the number of excitations in the subspace tends to zero and the entropy tends to a minimum.

It is obvious that the number of microstates increases with an increase in the volume of the subspace $M_{d_2}$. The inclusion of states with different four-momenta in the Minkowski space $M_4$ substantially enhances the effect.

Let us increase the dimension of the compact space with a large volume and consider the manifold of type(\ref{MD}), where $M_{D_1}=M_4\otimes M_{d_2}, M_{d_2}= S_2 , M_{d_1}= S_1$. It is known that the excitation mass in $S_2$ is $m(l)=\sqrt{l(l+1)}/r_2$, and the degeneracy factor is $2l+1$. The decay of excitation of the space $S_1$ into two excited states of the space $S_2$ also proceeds with energy conservation: $m_1 = m(l)+m(l')$. It is easy to see that, with
number of allowable energy states with zero momentum is also proportional to the volume of the large
space: $\Omega_2 \simeq \beta^2 =r_2 ^2 /r_1 ^2$.

\textbf{\textit{Therefore, if quantum fluctuations generate a space of the type $M_a \otimes M_b$, the entropy flow is directed toward the subspace with a larger size. The geometry of the subspace with a smaller size tends to a maximally symmetric geometry compatible with its topology. The existence of gauge symmetries in the main space appears to be a purely statistical effect.}}

\subsection{Rate of the Transition to the Symmetric State}

The entropy of the entire system consisting of two
subspaces (\ref{MD}), increases as a result of the transformation of particles (excitations) of the compact subspace $M_{d_1}$ to particles propagating in the subspace $M_{D_1}$. The compact extra space that, at the initial instant of time, has an arbitrary geometry acquires a maximally symmetric shape during the entropy transfer to the main
space. In this case, the entropy of the entire system (the extra and main spaces) increases and the entropy of the subsystem (the compact extra space) tends to a minimum. The situation resembles the third law of
thermodynamics, according to which the entropy of a
body tends to zero with a decrease in the thermostat
temperature.

The subspace $M_{D_1}$ in (\ref{MD}), should not necessarily have a direct-product structure. The main requirement for this subspace is the presence of a large number of energy levels that contribute to the statistical weight of the system at a fixed temperature. The Minkowski space $M_4$ itself exhibits this property in full measure.

Let us estimate the rate of ''symmetrization'' of the
extra space. Weak deviations of the geometry from an
equilibrium configuration can be interpreted as
excited states with the mass $m_1$,(see, for example, \cite{Antoniadis}). Since this is the only scale, it should be expected that the probability of decay will satisfy the relationship $\Gamma \sim m_1\sim
1/L_d$, where $L_d$ - is the characteristic size
of the extra space. Setting $L_d \leq 10^{-17}$cm, we find that the lifetime of the excited state is $t_1 \sim L_d \leq 10^{-27}$s. Therefore, the extra space transformed into the most symmetric state long before the onset of the primordial nucleosynthesis but, possibly, after completion of the inflationary stage. However, the states that correspond to the first excited level of the Kaluza - Klein tower with a lifetime of the order of $10^5$s or more is also considered \cite{Rizzo}. Under these conditions, the theory acquires the gauge invariance well after the nucleosynthesis stage, which creates new possibilities and problems. In order for a Kaluza - Klein particle to be stable, it is necessary
to make additional assumptions that complicate the
structure of the extra space \cite{Regis}.

\section{RELATION OF THE FUNDAMENTAL
CONSTANTS TO THE PROPERTIES
OF THE EXTRA SPACE}

Now, we discuss the problem associated with the
emergence of physical parameters in the modern Universe. In the sequential implementation of the
Kaluza - Klein idea, boson fields represent individual
metric components of the extra space. The only scale
possible in this situation is the length scale $\1$, I; however, the use of even this unit is complicated because quantum fluctuations in the spacetime foam continuously renormalize any parameters in an unpredictable manner. The situation becomes better as soon as a classical spacetime region arises due to the same quantum fluctuations. This is also accompanied by the appearance of independent stationary quantities that can be chosen as units of measurement, for example, the volume of the compact space $V_{d_1}$ and the vacuum energy density $U_m$. Only in this case, dimensional physical constants, including fundamental constants, can be fixed. A possible variant of this scenario will be considered below.

The action is chosen in the form (see, for example, \cite{Star,Nojiri,Sokolowski}),
\begin{eqnarray}\label{act0}
S = N_0 \int {d^D y\sqrt { - G} F(R;a_n)};\quad F(R;a_n) =
\sum\limits_n {a_n R^n },\quad a_1 =1.
\end{eqnarray}
The parameters $a_n , N_0$ take on specific values in the birth of this spatial region \cite{Rubin08} after the choice of the length unit. The appropriate choice of the parameters ${a_n}$ can provide the boundedness of the effective action from below \cite{BR06,our}. The dimension of the function $F(R)$ is conveniently chosen so that it coincides with the dimension of $R$; i.e., it is $\1^{-2}$. Then, the dimension
of $N_0$ is $\1^{2-D}$.

The metric of the space $M_{D}$ is written in the form \cite{Carroll,our}
\begin{eqnarray}\label{interval}
ds^{2}&=&G_{AB}dX^{A}dX^{B}=g_{ab}(x)dx^{a}dx^{b}-e^{2\beta(x)}\gamma_{ij}
(y)dy^{i}dy^{j} \\
&=& \rm{N}dt^2 - g_{\mu\nu}(x)dx^{\mu}dx^{\nu}-e^{2\beta(x)}\gamma_{ij}
(y)dy^{i}dy^{j}. \nonumber
\end{eqnarray}
Here $g_{ab}$ is the metric of the subspace $M_{4}=R\otimes M_{3}$ with the signature $(+---)$, $e^{2\beta(x)}$ is the radius of curvature of the compact subspace $M_{d}$, and $\gamma_{ij}(y)$ is its positive-definite metric. For a specified foliation of the space by spacelike surfaces, we can always choose the normal Gaussian coordinates, which is used in the last equality in expression(\ref{interval}). We assume that the coordinates $x,y$ have dimension $\1$, the time is measured in seconds, and the lapse function $ \rm{N}$ is an unknown parameter with the dimension of $(\1/s)^2$. The metrics $g_{ab}$ and $\gamma_{ij}$ are dimensionless.

According to \cite{BR06,our}, the topology and metric on the spacelike section $\Sigma_{f}$ of amplitude
(\ref{Amplitude}), are defined by
imposing the following conditions.

i) The topology of the space $M_D$ has the form of the direct product
\begin{equation}
M_D = M_4\otimes M_d,
\label{twosubsp}
\end{equation}
where ${d}$ is the dimension of the compact extra space.

ii) The curvature of the subspace $M_{d}$ satisfies the condition
\begin{equation}\label{Ineq0}
R_{4}(g_{ab})<< R_{d}(\gamma_{ij}).
\end{equation}

iii) As was shown above, the extra space with an
arbitrary geometry evolves to the space with a maximum number of Killing vectors that is possible for the topology under consideration. In this respect, in the set of subspaces $M_{d}$ we choose the maximally symmetric spaces with a constant curvature $R_{d}$, which is related to the curvature parameter $k$ in a typical manner
\begin{equation}\label{k}
R_{d}(\gamma_{ij})= \e^{-2\beta(x)}kd(d-1)\1^{-2}.
\end{equation}

Owing to the special form of the chosen metric (\ref{interval}), the following relationships hold true \cite{BR06,Carroll}
\bear                                   \label{R-decomp}
        R \eql R_{4} + \phi + f_{\rm der},
\nn
        \phi \eql kd(d-1) \e^{-2\beta(x)}\1^{-2}
\nn
      f_{\rm der} \eql 2d g^{\mu\nu}\nabla_{\mu}\nabla_{\nu}\beta
                + d(d+1) g^{\mu\nu}\d_{\mu}\beta \d_{\nu}\beta,
\ear where the field $\phi$ is defined by explicitly introducing its dimension $\1^{-2}$. The covariant derivative $\nabla$ acts in the
space $M_{4}$. The volume $\cV_{d}$ of the internal space of unit curvature depends on its geometry, because it is expressed through the intrinsic metric
\begin{equation}
\cV_d=\int d^{d}y\sqrt{\gamma}.
\end{equation}

In what follows, we will use the slow-change
approximation proposed in \cite{BR06}
\beq
|\phi| \gg |R_4|,\ |f_{der}|
\eeq
which is valid already at the onset of the inflationary stage. Then, by expanding the expression $F(R; a_n) = F(\phi + R_{4} + f_{der};a_n)$ в
into a Taylor series and integrating over the coordinates of extra dimensions, we obtain the expression
\bearr
     S \simeq \cV_d\, N_0                                \label{act3}
          \int \sqrt{^{4}g}\,d^{4}x\, e^{d\beta}
             [F'(\phi;a_n)R_4 + F(\phi;a_n) + F'(\phi;a_n) f_{der}],
\ear
which is typical of the scalar–tensor theory of gravity in the Jordan frame. By using the conformal transformation
\beq
\label{trans-g}
    g\mn \ \mapsto \tg\mn = |f(\phi)| g\mn, \cm
            f(\phi) =  e^{d\beta} F'(\phi;a_n),
\eeq
we change over to the Einstein frame
\bear\label{E-H_Action}
     S \eql \cV_d\, N_0 \int d^{4}x\, \sqrt{\tg}\, (\sign F') L,
\\
     L \eql R_{\1} + \Half K(\phi) (\d\phi)^2
                        - U(\phi) ,   \label{Lgen}
\\
     K(\phi) \eql                                \label{KE}
        \frac{1}{2\phi^2} \left[
            6\phi^2 \biggl(\frac{F(\phi;a_n)''}{F(\phi;a_n)'}\biggr)^2\!
            -2 d \phi \frac{F(\phi;a_n)''}{F(\phi;a_n)'} + \Half d (d + 2)\right],
\\
     U(\phi) \eql - (\sign F(\phi;a_n)')
        \left[\frac{|\phi|\cdot \1 ^2}{d (d -1)}\right]^{d/2}
                \frac{F(\phi;a_n)}{F'(\phi;a_n)^2 }                \label{VE}
\ear
where  $F(\phi;a_n)' = dF/d\phi$.

By assuming the existence of a minimum of potential (\ref{VE}) at $\phi =
\phi _m$ and using the slow-change approximation, action (\ref{E-H_Action}), in the vicinity of this minimum can be represented in the form
\begin{equation}\label{ActionI}
S \simeq N_0 \cV_d c_I \int {dt d^3 x_I \left[ {R_I  + \frac{1}
{2}K(\phi _m )(\partial_I \phi)^2 - U(\phi_m ) - \frac{1}{2}U''
(\phi _m )(\phi - \phi _m )^2 } \right] } ,
\end{equation}
where $$(\partial_I \phi)^2 = \left(\partial \phi /c_I \partial t
\right)^2 - \left( \partial \phi /\partial x_I  \right)^2$$. Here, we took into account that $\sqrt{g}=c_I$ in the four-dimensional Minkowski space. The index $\1$ designates the used unit of length, and $c_I$ is the speed of light in the chosen units.

The units of measurement are related by the expression
\begin{equation}\label{Units}
\1 =\alpha \cdot cm.
\end{equation}
где $\alpha$ is as yet unknown parameter. In relationship (\ref{ActionI}), we change over to the standard units of length
\begin{equation}\label{ActionIcm}
S = N_0 \cV_d c \int {dt d^3 x \left[R_4 + {\frac{1} {2}K(\phi _m
)(\partial \phi)^2 - \alpha^2 U(\phi_m ) - \alpha^2 \frac{1}{2}U''
(\phi _m )(\phi - \phi _m )^2 } \right] } .
\end{equation}
Here, $x_I =\alpha x$, $c_I =\alpha c$, $R_I =R_4 /\alpha^2$ and $(\partial_I \phi)^2=
(\partial \phi)^2 / \alpha^2$.

Since expression (\ref{act0}), is the low-energy limit of action (\ref{ActionIcm}), it should adequately describe purely gravitational phenomena. Moreover, expression (\ref{ActionIcm}) makes it possible to explain the origin of the inflaton potential and cosmological constant. The effective action (\ref{ActionIcm}) contains the initial parameters of the theory without using fundamental parameters, such as the Planck constant $\hbar$ and the gravitational constant $G$. However, from the practical and historical viewpoints, it is more convenient to explicitly introduce these parameters. In terms of the developed approach, this can be done by imposing the constraints
\begin{equation}\label{hbar}
N_0 \cV_d c =\frac{c^4}{16\pi G\hbar},\quad
\end{equation}
\begin{equation}\label{Lambda2}
\alpha ^2 U(\phi _m)=\frac{16\pi G}{c^4}\Lambda .
\end{equation}
Here $\Lambda$ is the observable vacuum energy (dark energy) density. It is important that the quantities, such as the volume of the extra space $\cV_d$ and the minimum value of the potential $U(\phi _m)$, acquire specific values only at $\phi = const$. If the field $\phi$ is identified with the inflaton, a nontrivial situation arises. The inflationary stage is completed before the scalar field (inflaton) reaches a potential minimum. Consequently, at the inflationary stage, when the Universe formed and the field $\phi$ varied
with time, the quantities $G$ and $\hbar$ should also depend on time. In this case, the theory of gravity is an effective theory valid at low energies (see also \cite{Volovik2}).

Let us introduce the definition
\begin{equation}\label{mass}
m_{\Phi}\equiv \alpha \sqrt{\frac{ U'' (\phi _m )}{K (\phi _m )}}
\end{equation}
and, under the assumption that $K(\phi _m )>0$, change the variables
$$\sqrt{\frac{c^4}{16\pi G} K(\phi _m )}(\phi  - \phi _m ) =\Phi .$$
As a result, we arrive at the conventional form of the
action for the scalar field $\Phi$ interacting with gravity; that is,
\begin{equation}\label{actinOrdin}
S = \frac{1}{\hbar}\int {dtd^3 x\left( {\frac{c^4}{16\pi G}R_4  + \frac{1}
{2}(\partial \Phi )^2  - \Lambda  - \frac{1}{2}m_\Phi ^2 \Phi ^2 } \right)} .
\end{equation}
Now, it becomes clear that the parameter $m_{\Phi}$ has the meaning of the scalar field mass.

Relationships (\ref{hbar}) and (\ref{Lambda2}) allow us to solve the problem of the separate emergence of the gravitational and Planck constants, which appear to be functions of the parameters of the theory $N_0 , \{ a_n\}$. The set of parameters $\{a_n\}$ is also implicitly included in the expressions for the quantities $\cV_d , U(\phi _m)$. In this paper, we do not consider the origin of the lapse function and, hence, the speed of light. By eliminating the insignificant parameter $\alpha$ from relationships(\ref{hbar}),
(\ref{Lambda2}) and (\ref{mass}), we obtain
\begin{eqnarray}
  \frac{U(\phi _m )K(\phi _m )}
{U''(\phi _m )} &=& \frac{16\pi G}
{c^4 }\frac{\Lambda }
{m_\Phi ^2 }  \label{connect1} \\
  N_0 \cV_d (\phi _m ) &=& \frac{c^3 }{16\pi G\hbar } .  \label{connect2}
\end{eqnarray}
Here, the right-hand sides contain measurable quantities, whereas the left-hand sides involve quantities
dependent on the initial parameters of the theory. At
energies on the Planck scale, when the field $\phi$ did not reach the potential minimum, the use of the modern values of the fundamental constants $G$ and $\hbar$ requires  care. Moreover, by assuming that the vacuum energies $\Lambda$ are different in different regions of the Universe \cite{varL,Rubin08,Dolgov}, the fundamental constants $G$ и $\hbar$ are also different according to expressions (\ref{connect1}), (\ref{connect2}).The discussion of this problem and references can be found
in \cite{Volovik}.

At the inflationary stage when $\phi = var$, the quantities $G$ and $\hbar$ also vary. The dependencies of these quantities on the field can be determined from (\ref{connect1}) and (\ref{connect2}) through the change $\phi _m \rightarrow \phi$, which holds
true for weak deviations of the field from the equilibrium position. As a result, we have
\begin{eqnarray}\label{Ghvar}
  G=G(\phi) &\simeq & \frac{c^4 m_\Phi ^2}{16\pi \Lambda} \frac{U(\phi )K(\phi )} {U''(\phi )} \\
  \hbar =\hbar (\phi ) &\simeq& \frac{c^3 }{16\pi G (\phi )\cV_d (\phi ) N_0 } .  \nonumber
\end{eqnarray}
The time dependence of the inflaton field $\phi$ for different inflationary models is well understood \cite{Lyth}, hence, the dynamics of the fundamental parameters can be described by choosing a specific inflationary model.

The factor of conversion between the units of
length $\alpha =\1 /cm$ can also be found by another method. Indeed, the characteristic size of the extra space is determined to be $\cV_{d}^{1/d}$ in units of $\1$. Moreover, by designating the size of the hypothetical extra space as $L_{d}$, expressed in terms of cm, we obtain $\alpha = L_{d}/\cV_{d}^{1/d}.$ Values of $L_{d} \leq 10^{-17}$ do not contradict experimental data. With due regard for expression (\ref{mass}), we find the constraint on the initial parameters
\begin{equation}\label{limit}
m_\Phi  \cV _d^{1/d} \sqrt {\frac{{K(\phi _m )}} {{U''(\phi
_m )}}} = L_d < 10^{ - 17}.
\end{equation}

Relationships (\ref{connect1}), (\ref{connect2}) and (\ref{limit}) allow us to determine the parameters of the theory $N_0 , a_n$ (see expressions (\ref{act0})) from the known constants $c,\hbar,G$ and the energy density $\Lambda$, the mass of the scalar field $m_{\Phi}$, and the volume of the extra space  $\cV_{d}$. The implementation of this program is a matter for the future. Actually, the fundamental constants are measured with a high accuracy, whereas the vacuum energy density is determined with a considerably lower accuracy. This is especially true in regard to the mass of the
hypothetical scalar field, even though this field is identified with the inflaton. The greatest uncertainty is associated with the volume of the extra space, which is
yet to be revealed.

\section{DISCUSSION}

In this paper, we have considered the problems of
the origin of the gauge symmetries and fundamental
parameters $\hbar$ and $G$. It has been shown that the modern values of the fundamental parameters and the
gauge invariance begin to be established in the inflationary period, i.e., at energy densities substantially lower than the Planck densities. The revealed dependence of the constants $\hbar , G$ on the geometry of the extra space and the initial parameters of the Lagrangian of the gravitational field with higher derivatives indicates that these parameters are not fundamental. According to string theory \cite{Landscape0,Landscape1}, and within the framework of the cascade birth of the Universe\cite{Rubin08}, the continuous formation of spatial domains generates domains characterized by different values of the effective parameters of the theory \cite{varL,Dolgov} and, in particular, by different values of the cosmological constant $\Lambda$. According to relationship (\ref{connect1}), this means that the gravitational and Plank constants are also subjected to spatial fluctuations.

The choice of the specific symmetry is a necessary
step in the construction of the theory. The use of gauge
symmetries makes it possible to effectively describe a
wide class of phenomena. It is worth noting that the
existence of symmetries themselves, as a rule, is postulated and that their origin is not refined. However, it is well known that, within the multidimensional
approach, the gauge symmetries are considered as a
consequence of the symmetries of the extra space. If
we accept the hypothesis that the extra space exists,
the problem is reduced to the elucidation of the factors
responsible for the existence of the Killing vectors in
this space. As has been shown in the present paper, the
extra space evolves to the maximally symmetric geometry, which is allowable by the corresponding topology due to the transition to the state with a minimum entropy. Meantime, the entropy of the entire system, including the extra and main spaces, increases. The conditions and rate of the symmetrization of the extra space and emergence of the gauge invariance in low-energy physics have been discussed. It has been demonstrated that, under certain conditions, this situation occurs only after the completion of the inflationary stage.

\section{ACKNOWLEDGMENTS}

The author is grateful to A. Berkov and D. Singleton for their interest expressed in this work and
S.V. Bolokhov, K.A. Bronnikov, V.D. Ivashchuk,
M.I. Kalinin, and V.N. Mel’nikov for helpful discussions.
This study was supported by the Russian Foundation for Basic Research (project no. 09-02-00677-a).

\end{document}